\documentclass[12pt]{article}
\usepackage{graphicx}

\newcommand\pubnumber{SNSN-323-63}
\newcommand\pubdate{\today}

\def\QM{School of Physics and Astronomy\\
Queen Mary, University of London, E1 4NS, United Kingdom}
\def\soton{School of Physics and Astronomy\\
University of Southampton, SO17 1BJ, United Kingdom}
\def\support{\footnote{On behalf of the T2K collaboration.}}

\def\Title#1{\begin{center} {\Large #1 } \end{center}}
\def\Author#1{\begin{center}{ \sc #1} \end{center}}
\def\Address#1{\begin{center}{ \it #1} \end{center}}

\newcommand\pubblock{\rightline{\begin{tabular}{l} \pubnumber\\
         \pubdate  \end{tabular}}}
\newenvironment{Abstract}{\begin{quotation}  }{\end{quotation}}
\newenvironment{Presented}{\begin{quotation} \begin{center} 
             PRESENTED AT\end{center}\bigskip 
      \begin{center}\begin{large}}{\end{large}\end{center} \end{quotation}}

\def\beq{\begin{equation}}
\def\eeq#1{\label{#1}\end{equation}}
\def\eeqn{\end{equation}}

\def\beqa{\begin{eqnarray}}
\def\eeqa#1{\label{#1}\end{eqnarray}}
\def\eeqan{\end{eqnarray}}

\let\bar=\overbar

\def\Dslash{\not{\hbox{\kern-4pt $D$}}}
\def\dslash{\not{\hbox{\kern-2pt $\del$}}}

\def\msb{{\bar{\ssstyle M \kern -1pt S}}}

\begin{document}
\begin{titlepage}
\pubblock

\vfill
\Title{Using the T2K near detector to study electron neutrino charged current quasi-elastic-like interactions on carbon}
\vfill
\Author{ Sophie King\support}
\Address{\QM}
\Address{\soton}
\vfill
\begin{Abstract}


In T2K $\nu_{\mu} \rightarrow \nu_{e}$ oscillation analyses, electron neutrinos are the signal at the far detector.  For the beam peak energy, $\sim 0.6 \,$GeV, charged current quasi-elastic (CCQE) neutrino interactions dominate.  Here we use the T2K near detector to measure electron neutrino CCQE-like interactions, which are defined by having no pions exit the nucleus:   $\nu_e \,$CC$\,0\pi$.  The selection, backgrounds and steps towards a cross section measurement on carbon are presented.

\end{Abstract}
\vfill
\begin{Presented}
NuPhys2015: Prospects in Neutrino Physics\\
London, UK,  December 16--18, 2015
\end{Presented}
\vfill
\end{titlepage}
\def\thefootnote{\fnsymbol{footnote}}
\setcounter{footnote}{0}
%

\section{Introduction:  T2K and the near detector}

T2K is a neutrino oscillation experiment that spans 295km across Japan \cite{t2kNIM}; the baseline is optimised to measure $\theta_{13}$ through electron neutrino ($\nu_e$) appearance in a predominantly muon neutrino ($\nu_{\mu}$) beam.  The neutrino beam peak energy, $\sim 0.6 \,$GeV, coincides with the first $\nu_e$ appearance probability maximum and enables T2K to exclude $\theta_{13} = 0$ with an impressive significance of 7.3$\sigma$ \cite{t2knueapp2014}.  

The far detector, Super-Kamiokande (SK), is a Cherenkov light detector situated at an off-axis angle of $2.5^{\circ}$ relative to the beam direction.  Positioned 280m from the source, along the same axis as SK, is the near detector, ND280.  This measures the flux, interaction rates and flavour content of the beam to constrain predictions at SK.  

\paragraph{Electron neutrinos at T2K \\}
For $\nu_{\mu} \rightarrow \nu_e$ oscillation searches, the signal at SK is $\nu_e$ and the biggest background comes from the intrinsic $\nu_e$ component of the beam itself.  The precision with which the $\nu_e$ cross-sections and intrinsic flux at SK are modelled consequently play a significant role in reducing the systematic errors of T2K oscillation results.  Previously the $\nu_e\,$CC-inclusive cross section was measured with ND280 data \cite{ben_nueCC_xs}.

Charged current quasi elastic (CCQE) interactions dominate at the T2K beam peak energy.  However, only particles that exit the nucleus may be detected, and since particles undergo final state interactions  in the nucleus, it is only possible to measure events that appear CCQE-like.  Pions, for example, may undergo scattering, absorption and charge exchange.  For this reason we define our signal in terms of particles exiting the nucleus, and focus here on $\nu_e $ charged current (CC) events with no pions in the final state,  $\nu_e \,$CC$\,0\pi$.  The remaining CC events are labelled $\nu_e \,$CC-other; these form a significant part of the background since the separation is limited by detector reconstruction efficiency.




\paragraph{The near detector\\} 

\begin{figure}[!ht]
\begin{center}
\begin{minipage}{0.40\textwidth}
\includegraphics[width=1\columnwidth]{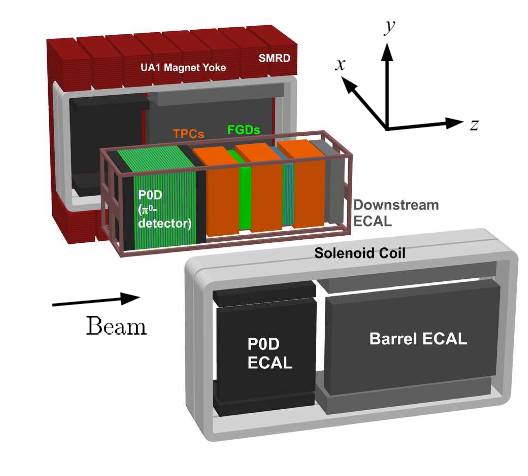}
\caption{Schematic of ND280}
\label{fig:nd280}
\end{minipage}
\hspace{24pt}
\begin{minipage}{0.36\textwidth}
\centering
\includegraphics[width=0.72\columnwidth]{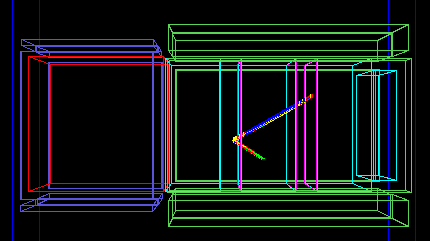}
\includegraphics[width=0.72\columnwidth]{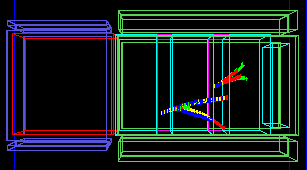}
\caption{ND280 display of simulated $\nu_e$ CC0$\pi$ (top) and $\nu_e$ CC-other (bottom) events}
\label{fig:events}
\end{minipage}
\end{center}
\end{figure}

ND280 comprises multiple sub-detectors, as depicted in Figure \ref{fig:nd280} where `downstream' (`upstream') is defined as the +z (-z) direction. Fine Grained scintillator Detectors (FGDs) provide an active target mass, and their size is optimised such that there is a good chance the lepton will travel through the adjacent Time Projection Chamber (TPC) and possibly the Electromagnetic Calorimeters (ECals).  One FGD has water layers between the scintillator to enable measurements on water, the target at SK.  The 3D reconstructed TPC tracks are used to calculate the momentum and charge of particles as they travel in the magnetic field.  Furthermore, the energy deposited as a function of distance gives excellent particle identification (PID) capabilities.  In the ECal, distribution of charge is used for PID where track-like (muon) and shower-like (electron) objects are distinguished.  FGDs/TPCs are numbered in the downstream direction, and upstream of these ND280 contains a $\pi^0$ detector (P$\emptyset$D).  The magnet is instrumented with a side muon range detector (SMRD).


\section{  $\nu_e \,$CC$\,0\pi$ event selection}
\label{sec:sigSel}
\begin{figure}[htb]
\centering
\includegraphics[width=0.384\linewidth]{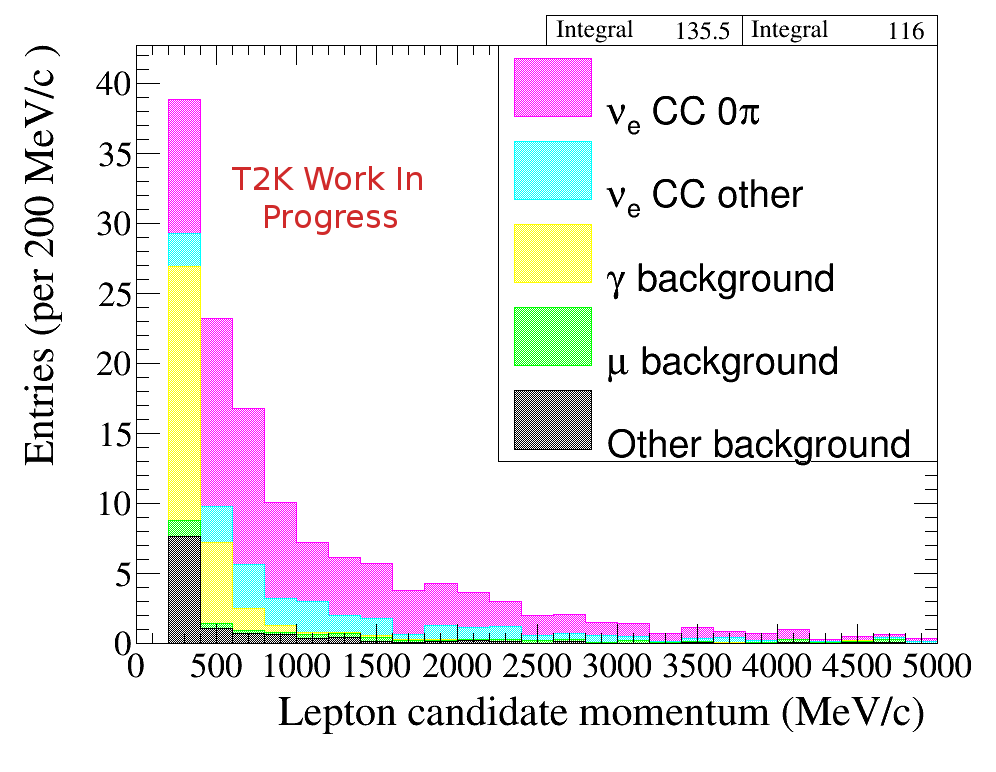}
\includegraphics[width=0.384\linewidth]{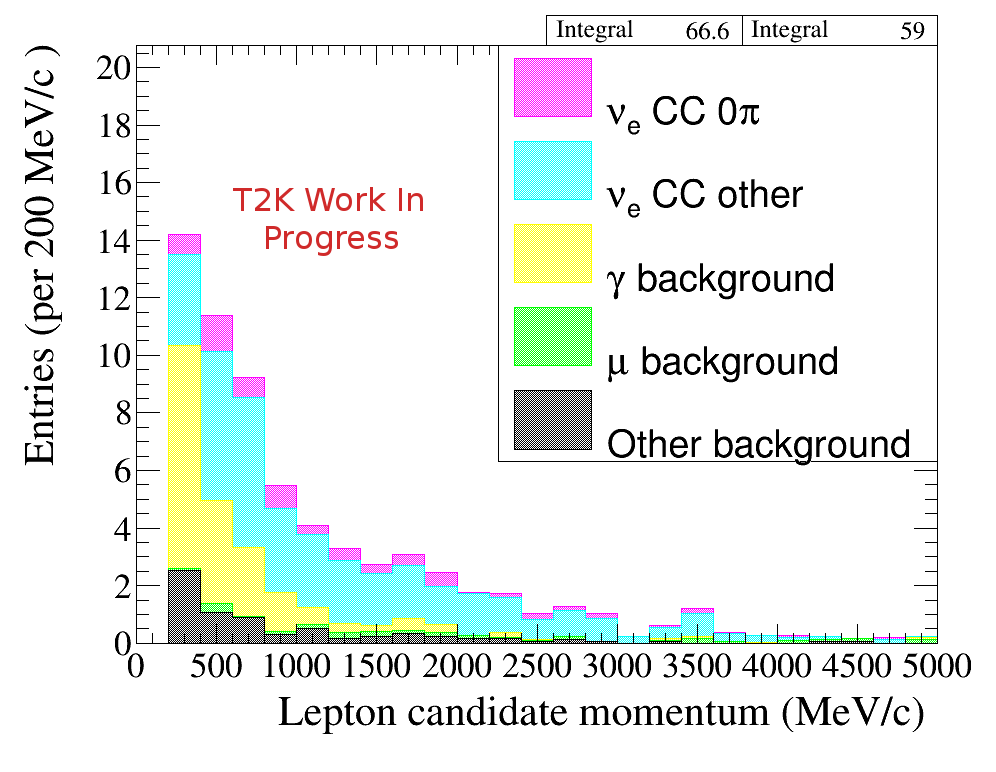}
\caption{Momentum distribution of $\nu_e \,$CC$\,0\pi$ (left) and $\nu_e \,$CC-other (right) samples.}
\label{fig:Sel}
\end{figure}

The $\nu_e \,$CC$\,0\pi$ selection process is detailed below and the final distribution is displayed as a function of momentum in the left of Figure \ref{fig:Sel}.  NEUT, a Monte Carlo event generator for neutrino interactions, is used to optimise the selection and produce plots.  
\vspace{4pt}
\\
\textbf{0) Event Quality} - Data quality and time compatibility checks are performed. 
\vspace{4pt}
\\
\textbf{ 1) Track selection} - The highest momentum negative track originating from FGD1 with a good quality TPC track is selected.  This is the lepton candidate.
\vspace{4pt}
\\
\textbf{2) PID} - To identify the selected track, the TPC uses (momentum dependent) energy deposited over the distance, and the ECal PID looks at charge distribution. 
\vspace{4pt}
\\
\textbf{3) Momentum cut} - Only events with a reconstructed momentum greater than $200\,$MeV are accepted. Below this the selection is dominated by $\gamma$-background.
\vspace{4pt}
\\
\textbf{4) Gamma Veto} - Events where the selected track is an $e^-$ from $\gamma \rightarrow e^- e^+ $ interactions are targeted; this is done by cutting on the invariant mass between the selected track and a second track that is positive and has an electron-like TPC track.
\vspace{4pt}
\\
\textbf{5) Upstream Vetoes}  -  Events with upstream activity in the P$\emptyset$D, ECal or TPC are removed; this indicates that the initial interaction  outside of the FGD.\vspace{4pt}
\\
\textbf{6) No Michel electrons} – Events with Michel electron candidates are rejected. 
\vspace{4pt}
\\
\textbf{7) Track multiplicity} - Events with additional FGD tracks are rejected.  In the case of only one extra track, events pass only if it forms a proton-like track in the TPC.

\section{Backgrounds and control samples}

The signal selection is estimated to be $\sim 53\,\%$ pure, with the two largest backgrounds due to $\nu_c$CC-other and the `$\gamma$-background' (due to $e^-$ coming from $\gamma \rightarrow e^- e^+ $).  These backgrounds, along with $\nu_{\mu}\,$CC$\,0\pi$, are constrained with dedicated control samples.

\paragraph{ $\nu_e \,$CC-other control sample \\}

Cuts $0-5$ in section \ref{sec:sigSel} are designed to select $\nu_e\,$CC interactions and therefore apply also to the $\nu_e \,$CC-other sample.  Events are then distinguished by the presence of Michel electrons or extra tracks. There is an upper bound on the $\nu_e$ CC-other track multiplicity to reflect events that may enter the signal selection due to detector reconstruction failure.  The resulting selection is shown in the right plot of Figure \ref{fig:Sel}.



\paragraph{$\gamma$-background control sample\\}

\begin{figure}[htb]
\centering
\includegraphics[width=0.384\linewidth]{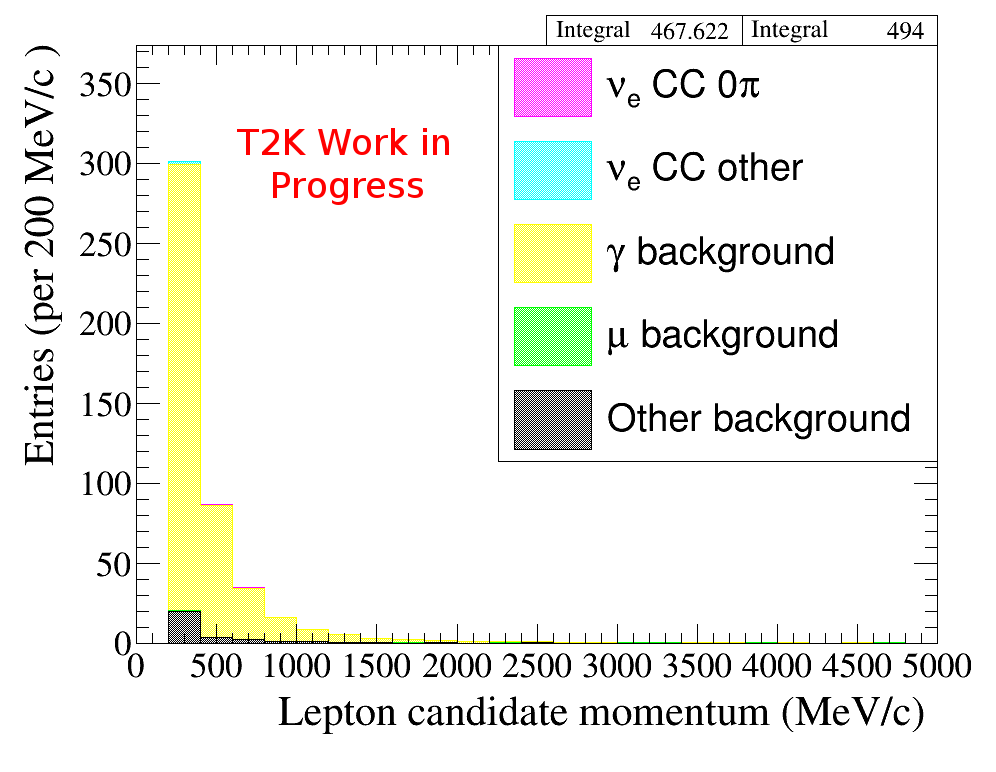}
\includegraphics[width=0.384\linewidth]{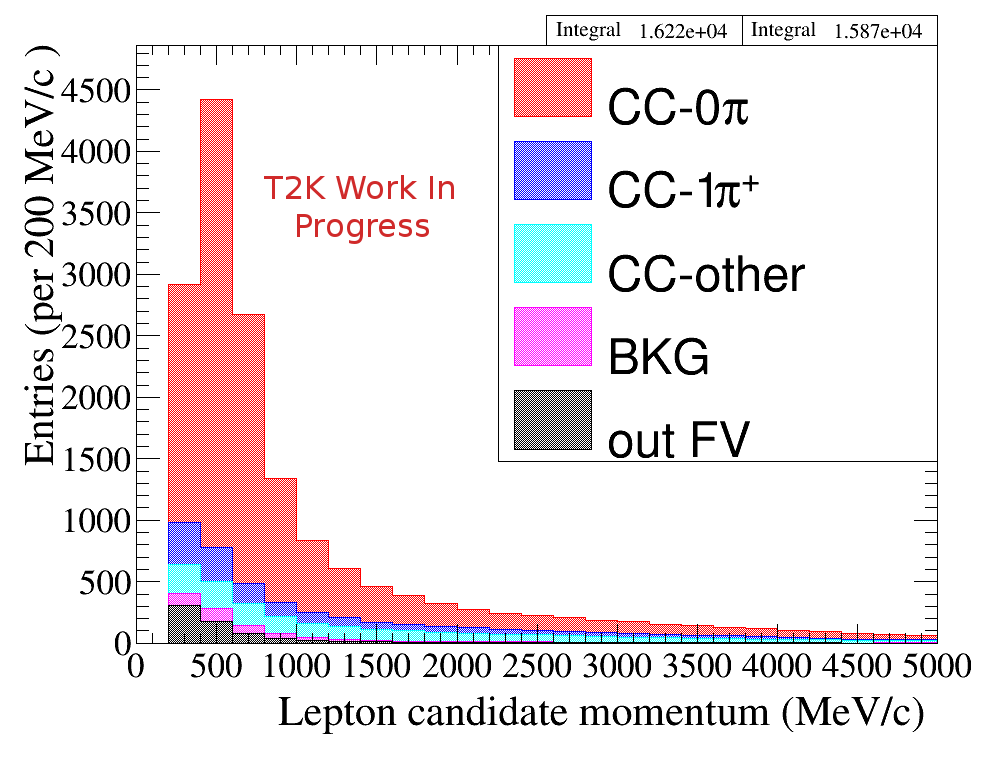}
\caption{The $\gamma$-background (left) and $\nu_{\mu}\,$CC$\,0\pi$ (right) control samples}
\label{fig:bkgs}
\end{figure}

Photons can travel through the detector unidentified and although the  $\gamma \rightarrow e^- e^+ $ conversion happens inside the FGD, it is quite possible, and indeed quite frequent, that the neutrino interaction vertex occurred outside it.  Consequently, $\gamma$ coming from outside the FGD have additional uncertainty compared to those that originate and convert in the FGD.  This is due to modelling of interactions on a wider variety of elements and possible miss-modelling of `dead' material (cables, joining material etc.).  A sample of $\gamma$-events are obtained using the inverse of requirements described in cut 4 of section \ref{sec:sigSel}; the resulting sample is displayed in the left plot of Figure \ref{fig:bkgs}.  Upstream activity is used to specifically target those coming from outside the FGD.

\paragraph{$\nu_{\mu} \,$CC$\,0\pi$ control sample\\}

Despite the good PID capabilities at ND280, the beam is $\sim 99\,\%$ pure in $\nu_{\mu}$ and sometimes a muon enters the signal selection.  To constrain this a $\sim 70\,\%$ pure sample of $\nu_{\mu} \,$CC$\,0\pi$ interactions are selected (right plot of Figure \ref{fig:bkgs} ) by identifying muons and events with no Michel electrons and no pion tracks.

\section{Towards a  $\nu_e \,$CC$\,0\pi$ cross section measurement}

This analysis is working towards a $\nu_e \,$CC$\,0\pi$ cross section measurement on carbon.  Control samples constrain the background, and Bayesian unfolding evaluates the  momentum, energy, angular and $Q^2$ dependence, in addition to a flux averaged result.

To account for the $\gamma$-background in the $\nu_e \,$CC-other sample, the normalisation fits are performed across the control samples simultaneously.  Initial fake data studies show that the small signal component in the $\nu_e \,$CC-other control sample has negligible effect in comparison to the statistical and systematic uncertainties, even when the signal prediction is significantly wrong.  Systematic and statistical uncertainties are expected to be $\sim 20\,\%$ and $\sim 15\,\%$ respectively.

\section{Summary}

The process for selecting $\nu_e \,$CC$\,0\pi$ events in ND280 is finalised, and the main backgrounds and sources of uncertainty identified.  Dedicated control samples constrain the backgrounds due to, $\nu_e$-CCother,  and  $\gamma$-background and $\nu_{\mu}\,$CC$\,0\pi$ and details of the cross section measurement on carbon are being finalised.



\end{document}